\documentclass[11pt,twoside]{article}

\usepackage{asp2006}
\usepackage{epsf}
\usepackage{psfig}
\usepackage{lscape}
\usepackage{natbib}
\begin{document}
\title{The Formation of Galaxy   Disks}   %%% Fill in title
\author{F.Governato$^1$, L.Mayer$^2$,  C.Brook$^1$}   
%%% Fill in author names
\affil{$^1$ University of Washington, Seattle, WA, $^2$ University of Zurich \&
 ETH}   
 %%% Fill in author affiliations

\begin{abstract} %%% Abstract to run on from here.

We present a new set of multi-million particle SPH simulations of the
  formation of disk dominated galaxies in a cosmological context. Some
  of these galaxies are higher resolution versions of the models
  already described in Governato et al (2007).  To correctly
  compare simulations with observations we create artificial images of
  our simulations and from them measure photometric Bulge to Disk
  (B/D) ratios and disk scale lengths.  We show how feedback and high
  force and mass resolution are necessary ingredients to form galaxies
  that have flatter rotation curves, larger I band disk scale lengths
  and smaller B/D ratios. A new simulated disk galaxy has an I-band
  disk scale length of 9.2 kpc and a B/D flux ratio of 0.64 (face on,
  dust reddened).
\end{abstract}

%%% MAIN BODY OF TEXT GOES HERE. CONSULT "INSTRUCTIONS FOR AUTHORS USING
%%% LATEX2E MARKUP", SECTIONS 2.3-2.6 FOR HELP WITH EQUATIONS, FIGURES,
%%% AND TABLES.
\section{The Simulation Campaign}

Here we present results from an ongoing project aimed at simulating a
representative sample of galaxies that form in a $\Lambda$CDM
cosmology. Simulations are run with GASOLINE \citep{wadsley04} and use the
``blastwave feedback'' described in \cite{stinson06}. Three of the four models
presented here use the same initial conditions as in Governato et al. (2007,
G07) and thus the galaxy halos have the same merging history and form inside
the same large scale structure. However, these new simulations use 8 times more
resolution elements and a 50\% smaller force softening than those presented in
G07.  To understand the numerical limitations of our work and to better compare
with the existing literature, we vary the number of resolution elements and the
feedback implementations.  In one case (named MW1) we run three models with the
number of resolution elements varying from a few tens of thousand to a few
million particles and force resolution from 1.2kpc to 0.3kpc.  To allow a
straightforward comparison between simulations and observations, the galaxies'
B/D ratios and disk scale lengths are measured using GALFIT on dust reddened,
face-on I and K band artificial images produced using {\sc SUNRISE}
\citep{jonsson06}. Recent works by different groups and using different
hydrodynamical codes have highlighted the role of feedback in creating galaxies
with a dominant stellar component supported by rotation
\citep{okamoto05,robertson04}. Significant improvements have been made in
reproducing a number of scaling properties of late type galaxies: namely the
age - circular velocity, the Tully Fisher, the stellar mass - metallicity
relations and the abundance of galaxy satellites
\citep{g07,brooks07,libeskind07}. However, a well known problem of most
galaxies formed in cosmological simulations is a massive, centrally
concentrated spheroidal component that creates unrealistically declining
rotation curves \citep{eke01,g07}.  The explanations proposed in the literature
invoke both physical and numerical arguments: catastrophic angular momentum
loss due to dynamical friction suffered by gas rich subhalos \citep{navarro96},
spurious gravitational torques between the disk and the surrounding halo of
dark matter and gas caused by the noisy mass distribution at low resolution
\citep{kaufmann07}, or artificial pressure gradients at the cold/hot gas
interface \citep{okamoto03}.
%\section{Separating the  Effects of Feedback and Resolution }   %%% Top level
\begin{figure}[!ht]
\plottwo{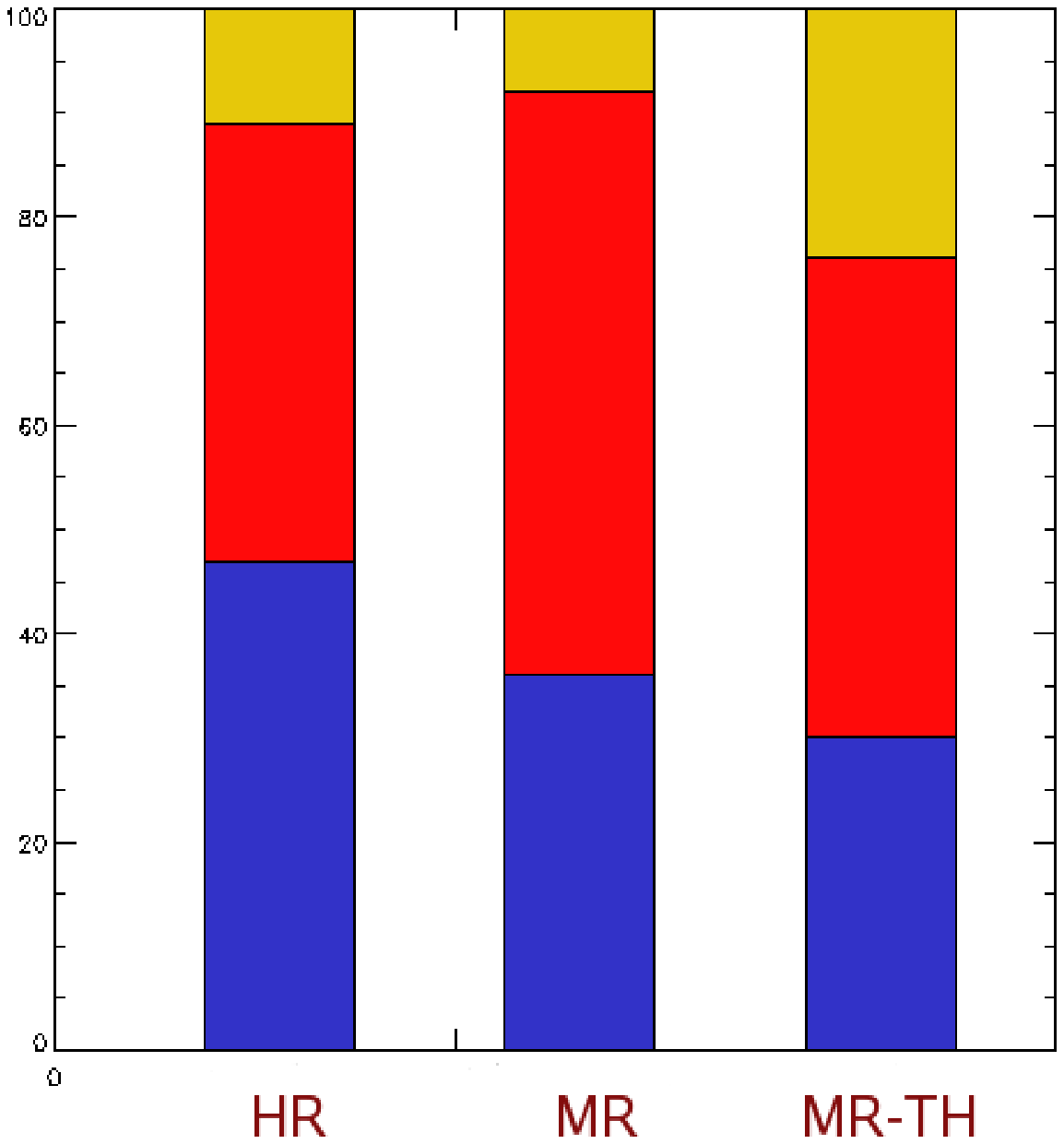}{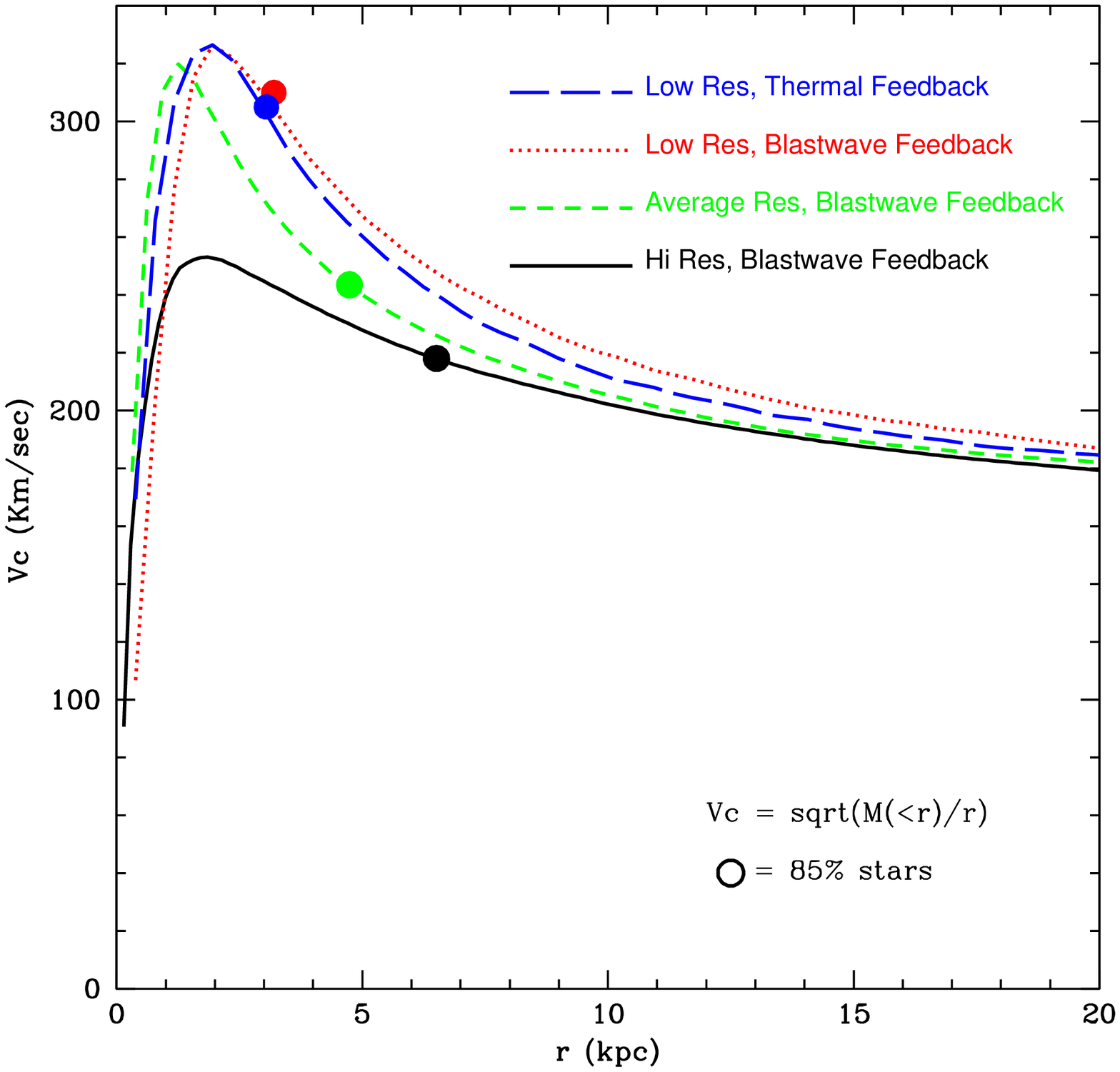}
\caption{Left: The kinematic disk, bulge and halo stellar mass fraction (blue,
  red and orange) in the MW1-HR, MW1-MR and MW1-MR-TH models. All runs use the
  ``blastwave'' feedback, but the one marked ``TH'', where ``thermal'' feedback
  is instead used. MR is the close equivalent of the MW1 galaxy described in
  G07. Histograms are normalized to 100\%. The total stellar masses differ by
  less than 10\%.  Right: The circular velocity profile from the total mass
  distribution (V$_c$ = (M/r)$^{1/2}$) for the same galaxy halo, run with
  different resolutions and feedback implementations. The circular dot marks
  the radius that contains 85\% of the stars. HR, MR and LR runs use a total of
  5 million, 600k and 70k DM, gas and star particles within the virial radius.}
\end{figure}
\begin{figure}[!ht]
\plottwo{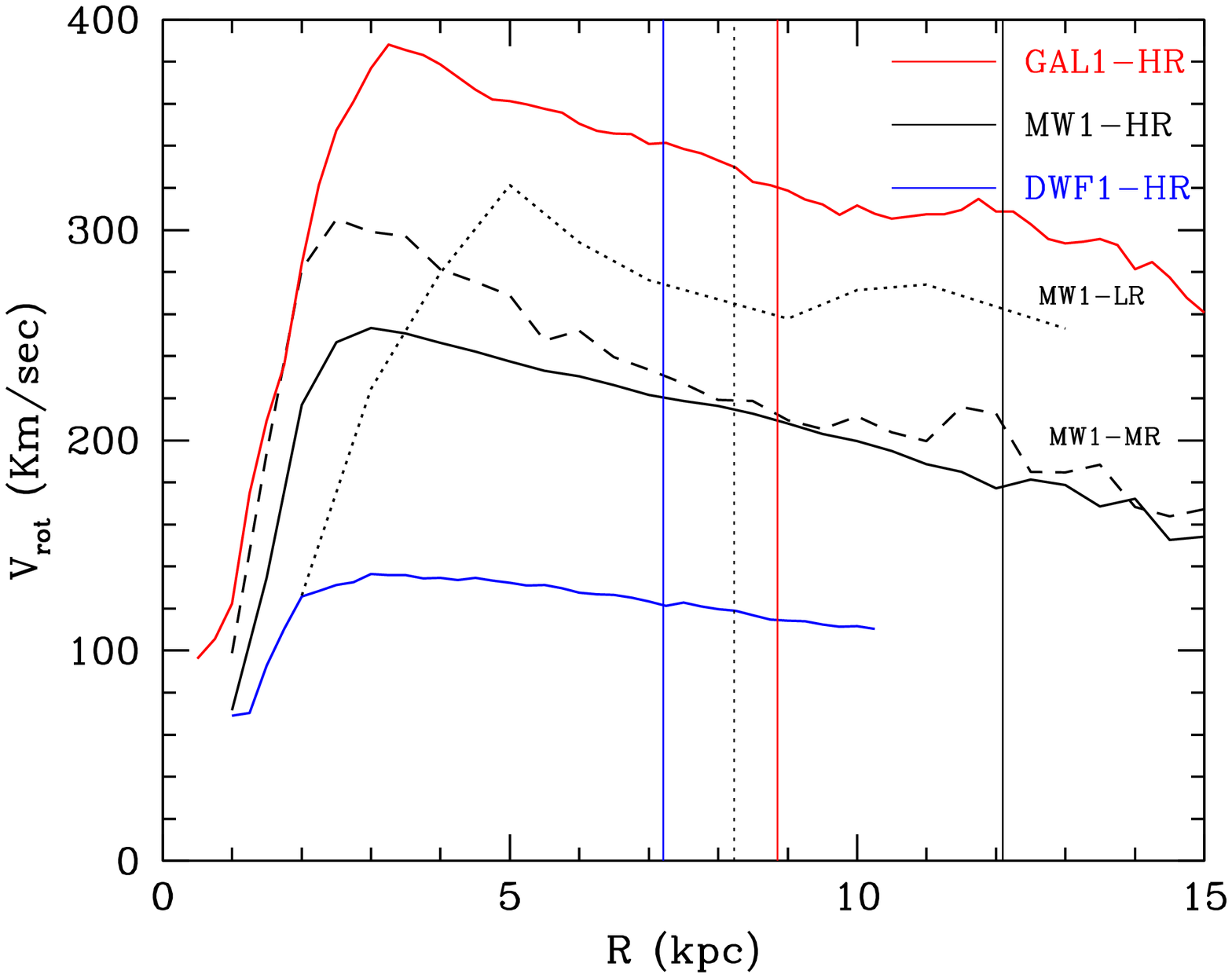}{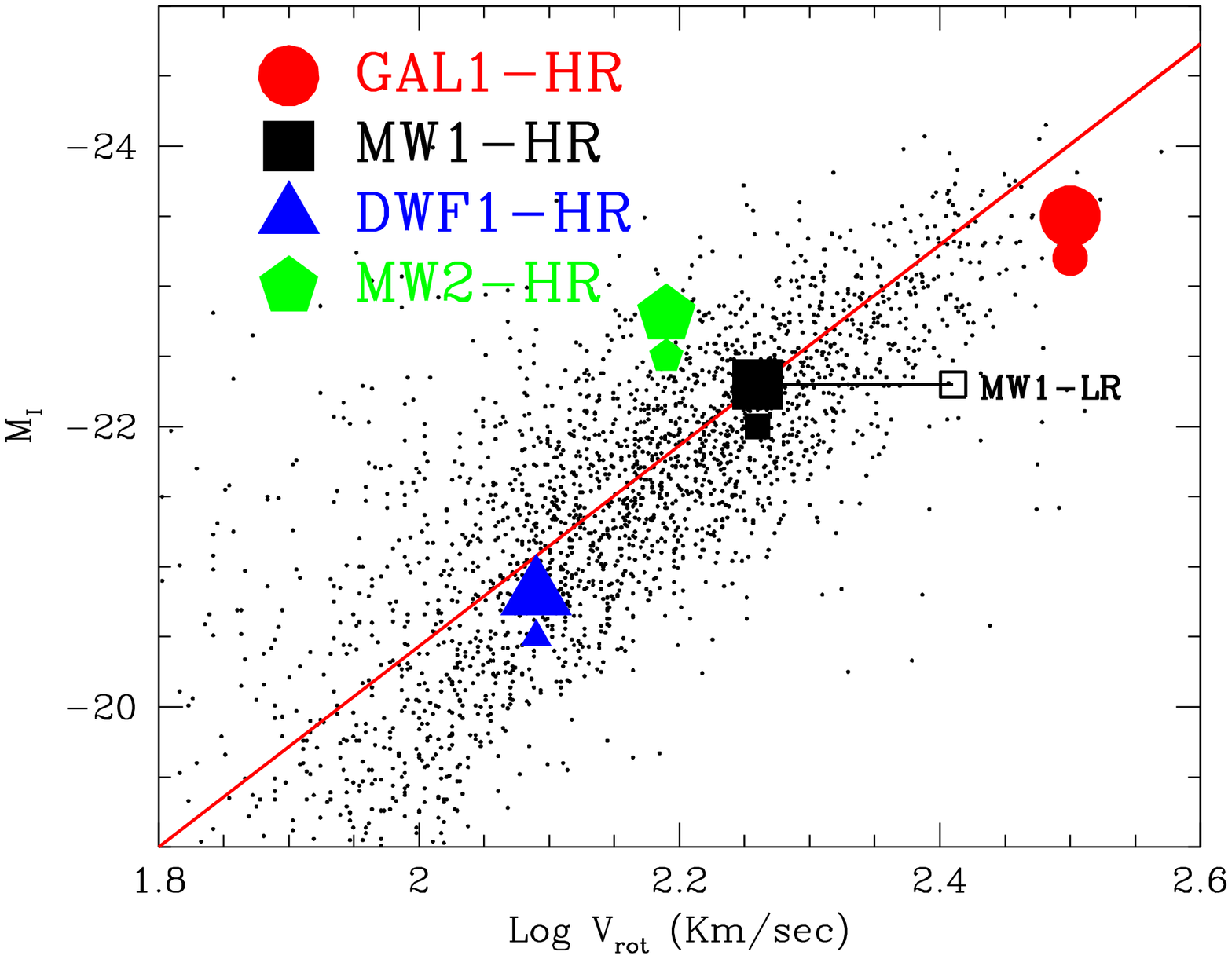}
\caption{Left: The  rotation curves for the higher
  resolution version equivalent of the models in G07. The dashed line is the
  curve for MW1 published in the paper, equivalent to the new ``MR'' model. The
  dotted line is the low resolution MW1-LR run. The vertical lines mark the
  radius at which V$_{rot}$ is measured: 3.5 I-band R$_d$s. This is a radius
  were the amount of included mass (DM, stars \& gas) is close to converging
  (Fig.1b) and gives more reliable results compared to measuring V$_{rot}$ at
  2.2 R$_d$ where numerical effects might still be present.  Right: the Tully
  Fisher relation for a sample of four high resolution runs vs. data in
  Giovanelli et al 97) Large filled dots: SSPs from Girardi et al
  (2000). Smaller filled dots: SSPs from Starburst99. Open square: MW1-LR. 
}
\end{figure}

\begin{table*}
\centering
\begin{tabular}{cccl}
\hline\hline
\noalign{\smallskip}
Run & I-band R$_d$ (kpc) & Kinem. B/D  & u -r \\
\hline
\noalign{\smallskip}
MW1-HR       & 3.5 & 0.83 (0.38) & 1.6   \\
MW1-MR       & 3.3 & 1.56 & 1.7  \\
MW1-MR-TH     & 2.1 &1.51 & 1.6 \\
MW1-LR        & 2.1  &1.35 &2.0  \\
\hline
\end{tabular}
\caption[Summary of the photometric properties of the MW1 runs] {Summary of the
photometric properties of the MW1 run obtained using SUNRISE and GALFIT. The
value in brackets is the photometric B-band B/D ratio for MW1-HR. MW1 has a
total mass of 10$^{12}$ M$\odot$}
\label{tbl-3}
\end{table*}

SN feedback is a promising solution as it decouples baryons from the dark
matter (Zavala et al., 2007).  However, \cite{naab07} recently presented results
from simulations that did not include feedback from SNe: the resulting circular
velocity profile of their simulated galaxies changed significantly with
increasing resolution, with the peak velocity dropping from 320 km/sec to 220
when the number of particles was increased from 40$^3$ to 200$^3$ (their
Fig.1). \cite{kaufmann07} used simulations of isolated galaxies to show that
cold, rotationally supported disks lose a large fraction of thier initial
angular momentum if the number of gas particles is below $10^6$.  The
literature presents encouraging evidence that both a physically motivated
description of feedback and a high number of resolution elements help to form
galaxies with at least some of the properties of the real ones.  However, an
analysis of a uniform sample of simulations with techniques {\it directly
comparable} to those used by observers (who measure the light rather than the
underlying mass distribution) has been lacking, making a comparison between
observations and theoretical models more open to interpretation.

\section{Results from Simulations}
To show the separate effects of feedback and resolution we focus on a
simulation of a Milky way like galaxy (total halo mass 10$^{12}$ M$\odot$). The
disk, halo and bulge components are first identified based on their kinematics
(Brook et al 08, in prep).  All runs adopted the ``blastwave'' feedback, but
those named ``TH'', that use the much less effective ``thermal'' feedback
\citep{katz92} which is a lower limit to the possible effects of feedback (the
two methods inject the same amount of energy into the ICM).  All runs include
the effects of a uniform cosmic UV background.  The kinematically defined
stellar disk component becomes more dominant as resolution increases or a
physically realistic description of feedback effects is adopted (Fig.1a). Bulge
masses decrease significantly in the high resolution run with feedback (HR vs
MR and \& MR-TH).  Feedback is also particularly effective in reducing the
fraction of halo stars, a consequence of a  smaller stellar mass in
stripped galaxy satellites (MR vs. MR-TH) (Brook et al 04).  Fig.1b shows the circular velocity
profile (V$_c$(r) = $\sqrt{M/R}$) for the same galaxy as resolution and
feedback are varied and shows results consistent with those in \cite{naab07}
and Kaufmann et al (2007): higher resolution models are less centrally
concentrated.  While this plot emphasizes the role of resolution, we have
verified that effect of feedback on the mass distribution is larger in less
massive halos with a shallower potential well, consistent with Zavala et al
(2007).
\begin{figure}
%\epsscale(1.85)
\plotfiddle{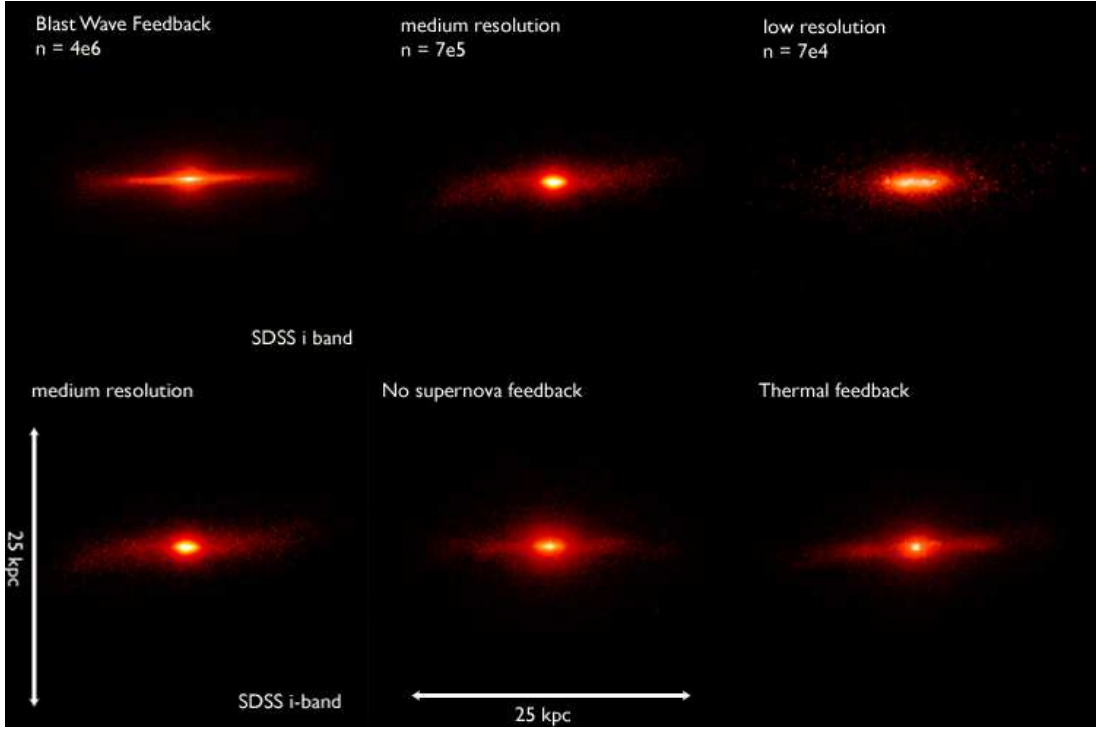}{2.75in}{-0}{55.}{55.}{-190}{0}
%\plotone{compare.eps}
\caption{{\sc Sunrise} Unreddened I-band images of the MW1 model. Upper panel:
  High to low resolution (with blastwave feedback). Lower panel: The medium
  resolution model with blastwave feedback, no SN feedback and ``Thermal''
  feedback. Galaxies with feedback and high resolution form disks with larger
  scale lengths and/or less concentrated bulges. The dust reddened,
  photometrically decomposed B/D ratio in B and I bands is lower than the one
  measured using the kinematics. }
\end{figure}
After a bulge/disk decomposition with GALFIT, photometric, dust reddened I-band
disks have larger scale lengths (R$_d$) and galaxies have bluer colors in
higher resolution runs that include the blastwave feedback (Table 1 and Fig.3). Fig.2a
shows the rotation curves from cold gas or young stars for the high resolution
version of the galaxies in G07. Compared with the lower resolution models in
G07 rotation curves are flatter, as the very central regions are less dense and
the disks dynamically colder in the outer parts.  However, they are still not
as flat as those of most real galaxies of similar mass. Fig.2b shows the
resulting I-band Tully Fisher relation. V$_{rot}$ is measured at 3.5
R$_d$. There is good agreement between data and the high resolution models,
confirming results in G07, while the low res MW1 model is significantly shifted
to the right compared to its high resolution equivalent and the observed
distribution.  We added to the sample a recently simulated large spiral galaxy
with an R$_d$ of 9.2, 9.2 and 9.1 kpc in the B,I and K bands respectively. Its
dust reddened I-band B/D is = 0.64.

%%% section head (remove "%" symbol)

%\subsection{}   %%% Second level section head (remove "%" symbol)
%\subsubsection{}   %%% Lowest level section head (remove "%" symbol)
   %%% Unnumbered top level section head (remove "%" symbol)
%\subsection*{}   %%% Unnumbered second level section head (remove "%" symbol)

%%% THE BIBLIOGRAPHY
%%%
%%% CONSULT SECTION 3 OF "INSTRUCTIONS FOR AUTHORS" FOR HOW TO USE NATBIB.
%%% AUTHORS ARE ENCOURAGED TO USE EITHER THE "THEBIBLIOGRAPY" ENVIRONMENT
%%% BY UNCOMMENTING (DELETING THE "%" SYMBOL) THE COMMANDS BELOW, OR BY
%%% USING THE BIBTEX ENVIRONMENT. TO FIND OUT WHICH IS APPLICABLE TO YOUR
%%% CONTRIBUTION, CONSULT THE VOLUME EDITORS FOR YOUR PROCEEDINGS.
%%%

\end{document}